\documentclass[twocolumn, pra,letterpaper,superscriptaddress ,amsfonts]{revtex4}
\usepackage{graphicx}

\usepackage{amsthm}
\usepackage{amsmath}
\usepackage{amsfonts}
\usepackage{labelfig}
\usepackage{amssymb}
\usepackage{bm}

\begin{document}

\bibliographystyle{apsrev}

\newcommand{\bra}[1]{\ensuremath{\left \langle #1 \right |}}
\newcommand{\ket}[1]{\ensuremath{\big| #1 \big \rangle}}

\newcommand{\braket}[2]{\ensuremath{\left \langle #1 \right | \left. #2 \right \rangle}}

\title{Photon Frequency Mode
Matching using Acousto-Optic Frequency Beam Splitters}
\author{Nick S. Jones}
\affiliation{Dept. of Mathematics, University of Bristol, University
Walk, Bristol, BS8 1TW, UK} \affiliation{Clarendon Laboratory,
University of Oxford, Parks Road, Oxford, OX1 3PU, UK}
\author{T.M. Stace}
\affiliation{DAMTP, University of Cambridge, Cambridge, CB3 0WA, UK}

\begin{abstract}
It is a difficult engineering task to create distinct solid state
single photon sources which nonetheless emit photons at the same
frequency. It is also hard to create entangled photon pairs from
quantum dots. In the spirit of quantum engineering we propose a
simple optical circuit which can, in the right circumstances, make
frequency distinguishable photons frequency indistinguishable. Our
circuit can supply a downstream solution to both problems, opening
up a large window of allowed frequency mismatches between physical
mechanisms. The only components used are spectrum analysers/prisms
and an Acousto-Optic Modulator. We also note that an Acousto-Optic
Modulator can  be used to obtain Hong-Ou-Mandel two photon
interference effects from the frequency distinguishable photons
generated by distinct sources.
%Engineering challenges can sometimes
%prevent both the creation of distinct solid state single photon
%sources which emit at the same frequency and the creation of
%entangled photon pairs from quantum dots.

\end{abstract}

\maketitle

\section{Introduction}

Frequency indistinguishable photons are very useful in protocols for
quantum computing and communication that exploit polarization, space
or time degrees of freedom. Unfortunately achieving this
indistinguishability can be practically very challenging, especially
when the photons are produced by solid state devices. Two photons
can be considered to be frequency distinguishable when their
bandwidths are much smaller than the difference in their average
frequencies. The practical challenges of tuning two distinct
physical processes so that they generate photons which are frequency
indistinguishable can sometimes be avoided. Instead, pairs of
photons can be rendered frequency indistinguishable {\em after} they
have been emitted. Variants of the circuit we propose can render
photons, which initially differ in frequencies by tens of megahertz
up to a few gigahertz, frequency indistinguishable  and we hope that
it will have broad uses across quantum optics.

Our circuit is  composed of only a pair of prisms and an
Acousto-Optic Modulator (AOM) which acts as a form of frequency beam
splitter \cite{huntington1,zhang,huntington2}. In \cite{huntington1}
the authors describe a radio frequency half-wave plate. They use an
highly asymmetric Mach-Zhender interferometer combined with a double
pass of an AOM and obtain a circuit which is very similar in effects
to the one described below (see also \cite{huntington3, huntington4}
for further work). Their aim was to demonstrate that qubits can be
defined and manipulated using sideband modes, whereas this paper
uses a Frequency Beam Splitter (FBS) to allow greater experimental
flexibility by allowing imperfect frequency mode matching to be
avoided.

 Having introduced the AOM and our circuit in Section
II, Section III shows that it can be useful in realizing
entanglement between separated sources (this is important for the
family of entangling schemes recently proposed for cluster state
computation \cite{rauss}). In Section IV we show how to use our
circuit to create entangled photon pairs from asymmetric single
quantum dots (such a resource has broad applications, from uses in
quantum cryptography \cite{Ekert91} to quantum computation
\cite{Browne05}).

\section{Acousto-Optic Modulators as Frequency Shifters}
In this section we review how AOMs are used as frequency shifters
and show how they can be used as the frequency analogue of a beam
splitter, i.e.  a frequency beam splitter (FBS). The AOM can be
viewed as a coupler of two electric fields via a phonon field
\cite{Steinberg}. Acoustic waves are used to generate periodic,
propagating, inhomogeneities in a crystal's refractive index and
these act as a moving diffraction grating, scattering incoming
light.

\begin{figure}[h!]
%\ShowGrid
%
 \SetLabels
 (0.22*0.8) $E_d(0)$ \\
 (0.22*0.17) $E_i(0)$ \\
 (0.79*0.8) $E_i(l)$ \\
 (0.79*0.17) $E_d(l)$ \\
  (0.5*-0.07) $\omega_i-\omega_d$ \\
\endSetLabels
\strut\AffixLabels{\includegraphics[angle = 0, width = 4.5cm,
 keepaspectratio=true]{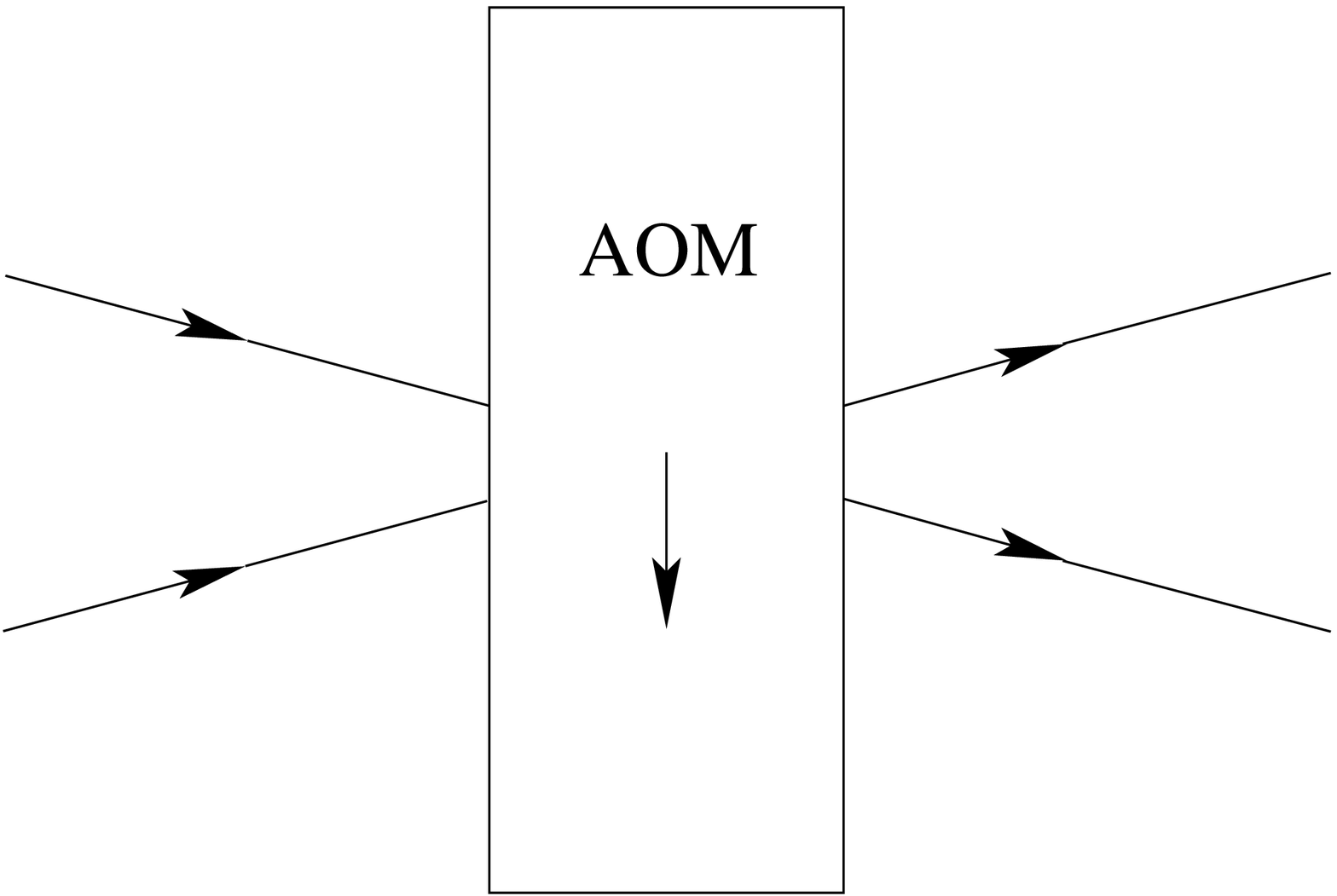}}
\caption{\label{fig1}  The classical model of the AOM as a coupler
of two fields. $E_d(0),E_i(0)$ are the field components in the
incident direction at the crystal input (at frequencies $\omega_d$
and $\omega_i$ respectively)
 and $E_d(l),E_i(l)$ the fields at the output. The arrow inside
the box indicates the direction in which the acoustic wave is
traveling and the crystal is modulated at the frequency difference
between the two input fields, $\omega_i-\omega_d$ . }
\end{figure}

The classical relationship between the fields at the input,
($E_i(0), E_d(0)$), and output, ($E_i(l), E_d(l)$), if both beams
travel a distance $l$ through the crystal is:
\begin{eqnarray}
&& E_i(l)=E_i(0)cos(\eta l) + iE_d(0)sin(\eta l), \label{equation1}\\
&& E_d(l)=E_d(0)cos(\eta l) + iE_i(0)sin(\eta l),
\label{equation2}
\end{eqnarray}
The constant $\eta$ has the following form:
$\eta=\frac{\omega}{2\sqrt{2} c}\sqrt{(\frac{n^6p^2}{\rho
v_s^3}I_{{acoustic}})}$, here $I_{{acoustic}}$ is the acoustic
intensity of the sound waves applied to the crystal, $n$ is its
refractive index, $v_s$ its speed of sound, $p$  its photo-elastic
constant and $\rho$ its density \cite{yariv}.  Taking $\omega_i$ as
the frequency of $E_i(0)$ and $\omega_d$ that of $E_d(0)$ and
assuming that $\omega_i-\omega_d\ll\omega_d$  we then define the
$\omega$ in the expression for $\eta$ as  $\omega = \omega_i \simeq
\omega_d$, see Fig. 1. Considering only the forward propagating
field components and quantizing these, the input and output modes
are coupled as below:
\begin{eqnarray}
&& b_i=a_icos(\eta l)+ ia_dsin(\eta l),\\
&& b_d=a_dcos(\eta l) + ia_isin(\eta l),
\end{eqnarray}
where $a_i$ is the annihilation operator for modes at the input
frequency $\omega_i$ and $a_d$ for those at frequency $\omega_d$ and
the frequency of the modulation of the crystal is
$\omega_i-\omega_d$.
%\begin{figure}[h!]
%\includegraphics[angle = 0, width = 5cm, height =3.5cm]{nick1.eps}
%\caption{\label{fig2}  A frequency beam-splitter, FBS.
%Prisms/spectrometers which can resolve energy differences of the
%order of meV are available.}
%\end{figure}

\begin{figure}[h!]
%\ShowGrid
%
\SetLabels
\L (0.41*0.65) $\omega_d$ \\
\L (0.41*0.30) $\omega_i$ \\
\L (0.80*0.65) $\omega_i$ \\
\L (0.80*0.30) $\omega_d$ \\
\L (0.56*-0.1) $\omega_i-\omega_d$ \\
\endSetLabels
\strut\AffixLabels{\includegraphics[angle = 0, width = 8cm,
 keepaspectratio=true]{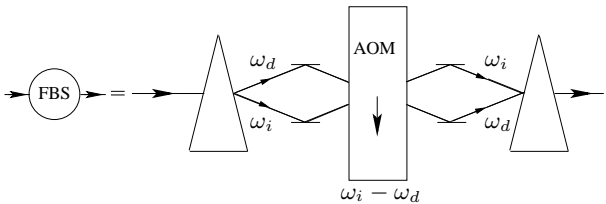}}
\caption{\label{fig2}  A frequency beam splitter, FBS. The AOM is
modulated at the difference of the two frequencies $\omega_i$ and
$\omega_d$.}
\end{figure}
Consider the circuit in Fig. 2. %which resembles a spatial
%beam-splitter in its action.
 The first prism  splits the state by
its frequency, the AOM rotates  between the two frequencies and the
second prism recombines the state. It effects the map:
\begin{eqnarray}
&&\alpha \ket{\omega_i} +\beta\ket{\omega_d} \rightarrow (\alpha
cos(\eta l) + i\beta sin(\eta l))\ket{\omega_i} \nonumber
\\
&&\;\;\;\qquad\qquad\qquad+ (i\alpha sin(\eta l)+ \beta cos(\eta
l))\ket{\omega_d}.
\end{eqnarray}
Consider modulating the AOM at $\omega_i-\omega_d$ and arranging the
apparatus so that $\eta l=\pi/4$ (giving a 50$\%$ conversion
efficiency). In this case the circuit performs the maps
$\ket{\omega_i}\rightarrow \frac{1}{\sqrt{2}}\ket{\omega_i}+i
\frac{1}{\sqrt{2}}\ket{\omega_d}$ and $\ket{\omega_d} \rightarrow  i
\frac{1}{\sqrt{2}}\ket{\omega_i}+ \frac{1}{\sqrt{2}}\ket{\omega_d}$.
%and the resulting states (unlike $\ket{\omega_i}$ and
%$\ket{\omega_d}$) cannot be distinguished by frequency measurements
%alone.
This is similar to a spatial beam splitter in its action. It
should be noted that $\eta$ depends on the frequency of the incident
light and thus bandwidth effects could make the action of the above
FBS more complicated. In Section IV we argue that  this will not
prove a practical issue for relevant photon bandwidths. Also note
that AOM's can only be modulated at frequencies up to a few GHz
\cite{brimrose}. In practice, the prims we describe would need to be
very good spectrometers. Devices which can resolve frequency
differences, $\omega_i-\omega_d$, of less than a GHz are just on the
edge of what is currently experimentally achievable.  By using the
FBS in slightly modified ways the schemes presented below avoid the
need for such high resolution spectrometers.

%appear to be on the edge of experimental

%biggest ghz implies v good prisms on edge of attainable

%cut paste mention to the edge of expt resolution from earlier

%note that this problem is avoided in the schemes presented - because
%the prisms are either removed or split widely separated frequencies

%correct the caption

In what follows we provide two applications for variants of the
above circuit. The first is in entangling emitters placed in
cavities, the second is in entangling correlated photon pairs from
quantum dots.

\section{Single photon use: entangling ions in cavities}
A number of schemes  exist for entangling pairs of atoms or dots
which have been placed in spatially separated cavities/traps \cite{
cabrillo, bose, protsenko, barretts,Lim2004, duan, Stace05}. To be
concrete we will consider the protocol of Barrett and Kok
\cite{barretts}; they have a technique for generating the highly
entangled cluster states needed in one-way quantum computation
\cite{rauss}. They suggest using the apparatus in Fig 3a to entangle
two photon sources. Either a single photon comes from source $1$ or
$2$. Simultaneous emission events from $1$ and $2$ are discarded. If
the photons from 1 and 2 are at the same frequency, the presence of
the beam splitter ensures that, when one of the detectors clicks,
the experimenter cannot determine whether the photon came from
cavity 1 or 2. For solid state single photon sources
\cite{yamamotolatest} it is very difficult to fix the frequencies of
the two sources to be the same, as is required in Fig. 3a; the
photons from directions 1 and 2 can often be frequency
distinguished. Though the authors suggest filtering out unmatched
photons, this leads to an appreciable reduction in the efficiency
with which their scheme makes large cluster states.

\begin{figure}[h!]
%\ShowGrid
%
\SetLabels
\L (0.75*0.46) $\omega_1-\omega_2$ \\
\L (0.775*0.41) $100\%$ \\
\L (0.33*-0.01) $\omega_1-\omega_2$ \\
\L (0.37*-0.055) $50\%$ \\
\L (0.24*.23) $\omega_2$ \\
\L (0.24*.10) $\omega_1$ \\
\L (0.51*.23) $\omega_1$ \\
\L (0.51*.10) $\omega_2$ \\
\endSetLabels
\strut\AffixLabels{\includegraphics[angle = 0, width = 7cm,
 keepaspectratio=true]{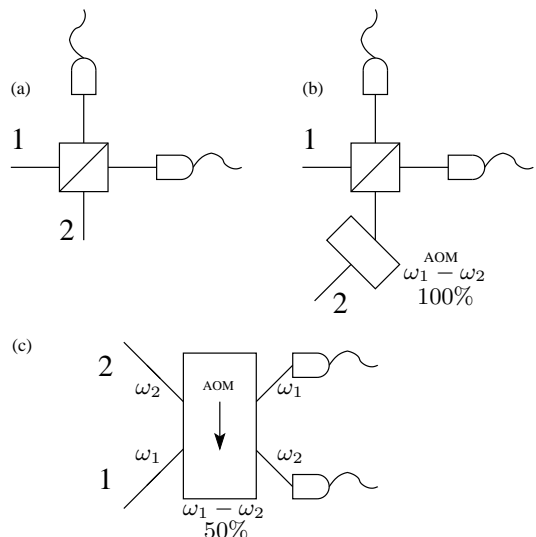}}
\caption{\label{fig3}  a) Photons either from cavity 1 at frequency
$\omega_1$ or cavity 2 at frequency $\omega_2$ are incident on
non-polarizing beam splitters. b) Photons from cavity 2 are
frequency up-shifted with ideally 100$\%$ efficiency so that they
have the same frequency as photons from cavity 1. c) In this scheme
the AOM need only frequency shift half the light - 50$\%$
efficiency}
\end{figure}

A first solution is to use an AOM on one of the photons to frequency
match the pair as in Fig. 3b. It is, however, difficult to perform
this frequency shift with 100$\%$ efficiency - normally some light
is unshifted (highly efficient AOM's also absorb more light - so
100$\%$ efficient devices with low absorptivity are difficult to
fabricate). Fig. 3c provides an alternative approach. Conceptually
this apparatus moves the challenge of tuning distinct components to
the easier task of tuning the apparatus of detection. %The dichroic
%mirror (or equivalent) is designed to transmit the photons at
%frequency $\omega_1$ from cavity 1 and reflect those at $\omega_2$
%from cavity 2. It is the frequency distinguishability of the two
%photons that enables this, otherwise impossible,  mode matching.
The photon passes through
 a 50$\%$ efficient frequency beam splitter at modulation
frequency $\omega_1-\omega_2$. From a click at either detector, the
experimenter will be unable to deduce the  cavity from which the
photon came. If the cavities differ in their frequency of emissions
by up to a few gigahertz the circuit can still allow them to become
entangled. We have increased the window of suitable emitters.
Depending on the size of the frequency shift, an AOM with 50$\%$
diffraction efficiency should be feasible with current components
\cite{brimrose}.

Though Fig 3c was described in the context of single photons, it
also has a two photon role. Experiments investigating the
Hong-Ou-Mandel dip \cite{HOM} for successive photons from the same
single photon solid-state source have already been studied
\cite{yamamotodist}. The scheme in Fig. 3c could be used to detect a
dip between photons from {\em distinct} sources when the sources are
not frequency matched.

\section{Two Photon use: Rectifying polarization entanglement in biexciton emission}
In this section we provide a use for frequency beam splitters in the
generation of entangled photon pairs. Quantum dots can emit
correlated photon pairs by a cascade of photon emission from a
biexciton via an exciton \cite{ben00}. Asymmetries in the dot can
prevent the photon pairs from being entangled. A symmetric dot might
generate the two photon state
\begin{equation}\label{sps1}
\ket{\phi}=
\frac{1}{\sqrt{2}}(\ket{x,\omega_A;x,\omega_B}+e^{i\nu}\ket{y,\omega_A;y,\omega_B}),
\end{equation}
where $\ket{x,\omega_A}$ is a photon linearly polarized in the
$x$-direction at frequency $\omega_A$ (similarly for
$\ket{x,\omega_B}$) and $\ket{y,\omega_A}$ is a $y$-polarized photon
at frequency $\omega_A$ ($\ket{y,\omega_B}$ similarly) and $\nu$ is
a phase  \cite{note1}. In practice asymmetries are hard to avoid
when constructing the dots \cite{kulakovskii} and the following
state might be expected instead
\begin{equation}
\ket{\phi'}=
\frac{1}{\sqrt{2}}(\ket{x,\omega_1;x,\omega_2}+e^{i\nu}\ket{y,\omega_3;y,\omega_4}),
\end{equation}
where $\omega_3>\omega_1>\omega_2>\omega_4$ and $\omega_3=\omega_1 +
\Delta=\omega_2 + \Delta+\xi=\omega_4 + 2\Delta+ \xi$. We have
$\Delta=(\omega_3-\omega_1)=(\omega_2-\omega_4)$ as the doublet
splitting due to dot asymmetry and $\xi=\omega_1-\omega_2$ as the
biexciton shift. Since $\ket{x,\omega_1}$ and $\ket{y,\omega_3}$ are
frequency distinguishable (and also $\ket{x,\omega_2}$ and
$\ket{y,\omega_4}$) no polarization entanglement will be found
\cite{Stace}. Each photon will also have a certain narrow bandwidth
rather than being strictly monochromatic. The lack of coherence
detected in emissions from such dots \cite{Santori,ste02}
 has been understood as indicating that there is no appreciable frequency overlap between
 each of the photons emitted at different frequencies (their bandwidth
 is less than the doublet splitting). Rather than tuning the dots to eliminate asymmetries
 or using cavities to control the emitted
frequencies \cite{Stace} our circuit offers a downstream solution
(see Fig. 4a,b) with general applicability. In Fig. 4a  the AOM is
modulated at $\omega_3-\omega_1$ and  performs the map:
\begin{eqnarray}
&&\ket{\phi'}\rightarrow
\frac{1}{\sqrt{2}}(\ket{x,\omega_1;x,\omega_2}+e^{i\kappa}(i\sin(\eta
l)\ket{y,\omega_1}+ \nonumber\\ &&\cos(\eta l) \ket{y,
\omega_3})(\cos(\eta l) \ket{y, \omega_4} + i\sin(\eta l)\ket{y,
\omega_2}).
\end{eqnarray}
If the AOM is 100$\%$ efficient at the desired frequency ($\eta
l=\pi/2$) then the polarization entangled state
$\frac{1}{\sqrt{2}}(\ket{x,\omega_1;x,\omega_2}+e^{i(\kappa+\pi)}\ket{y,\omega_1;y,\omega_2})$
is obtained (compare with Eq. (\ref{sps1})).

%[angle = 0, width = 10cm, keepaspectratio=true]

\begin{figure}[h!]
%\ShowGrid
%
\SetLabels
\L (0.42*0.85) $X$ \\
\L (0.29*0.575) $Y$ \\
\L (0.425*0.47) $\Delta$ \\
\L (0.575*-0.06) $\Delta$ \\
\L (0.295*0.235) $\omega_4$ \\
\L (0.295*0.105) $\omega_3$ \\
\L (0.84*0.235) $\omega_2$ \\
\L (0.84*0.105) $\omega_1$ \\
\endSetLabels
\strut\AffixLabels{\includegraphics[angle = 0, width = 8cm,
 keepaspectratio=true]{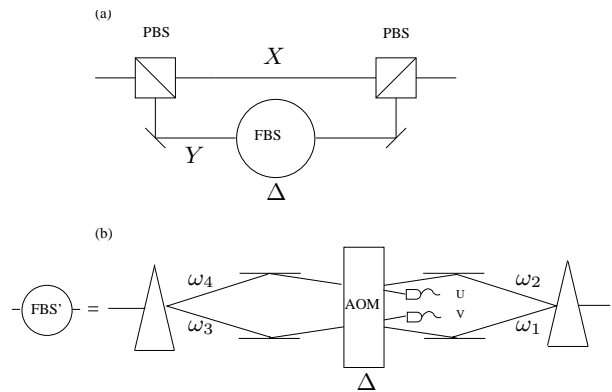}} \caption{\label{fig4}a) To demonstrate
  two-photon interference using the Mach-Zehnder interferometer, X polarized photons
take the top path and Y polarized photons are reflected and pass
through the FBS defined in Fig 2 (modulated at the frequency
$\Delta$). Here PBS is a polarizing beam splitter. b) Assuming that
AOM's do not frequency shift all incident light we propose this
circuit, FBS', as a substitute for the FBS in a). Here photon
detectors are placed at points U and V and these detect all
frequency unshifted photons. Unlike the situation in Fig. 2 and Fig.
3c, there is  no longer a mixing of frequency shifted and unshifted
photons; there is a sense in which the AOM is being used as two
separate devices. The AOM is thus not modulated at the frequency
difference of the input beams, as was the case in Fig. 2 and Fig
3c.}
\end{figure}

Depending on the dot's specific asymmetries a frequency difference
$\Delta$ of the order of 800MHz might be obtainable. %\cite{shefield}
There exist AOM's that operate at this frequency  (shifting by
around 5$\mu$eV) - though they cannot shift all incident light to
different frequencies (typically
 efficiencies are $<$80$\%$ \cite{brimrose}). It is thus
practically difficult to have $\eta l= \pi/2$. If one wants a
maximally entangled state, $\ket{\phi}$, but cannot frequency shift
all light or perform complicated multiple or single copy
purifications we propose the substitution of the circuit FBS' (Fig.
4b) for the FBS in Fig. 4a. In FBS' the two photons are being
addressed individually and do not cross in the AOM, so though the
circuit looks like Fig. 2 it is actually slightly different. One
could, alternatively, use two distinct modulators operating at the
same frequency on each arm of FBS'. Each frequency shift fails with
probability $(1-\alpha)$ and then the unshifted photon can be
detected. For perfect detectors in FBS', if neither clicks the
desired frequency shift has been obtained (this occurs with
probability $\alpha^2$) conversely any clicks herald a failure. In
practice single photon detectors produce both false positives and
negatives and are not essential here. Photon loss at U, V, will mean
that any two photon experiments will fail with probability
$1-\alpha^2$ but with current technology two photon events are rare
so a factor of $\alpha^2$ should not significantly change count
rates. AOM's typically absorb $<$0.05$\%$ of incident radiation so
the circuit will produce maximally polarization entangled photons
with probability $\gtrsim 0.95^2\alpha^2$ ($\alpha<0.8$).

\bigskip

The quantity $\eta$ is frequency dependent. This means that photons
at different frequencies are shifted in differing proportions. It is
assumed in the derivation of Eqs. (\ref{equation1},\ref{equation2})
that the two input frequencies are close (that the crystal
modulation frequency is small compared to the light frequency) but
this must be checked in this case. The biexciton shift, $\Delta$, is
$\sim 1\times10^9$Hz (by contrast $\xi\sim 1\times10^{11}$Hz) and
typical optical photon frequencies are $\sim 1\times10^{15}$Hz. The
proportion of light of frequency $\omega$ which is frequency shifted
is $\sin^2(\omega R)$ where $R$ is a constant depending on the
crystal properties and the intensity and frequency of modulation
\cite{yariv}. For GaP crystals of 1mm thickness modulated with
intensity 1$\times 10^{-6}\; W/m^2$, $R$ is $\sim 1\times
10^{-15}\;s$. The ratio of frequency shifts for photons at
frequencies $\omega$ and $\omega+\delta$, $\delta$ small, is
$\sin^2((\omega + \delta)R)/\sin^2(\omega R)$ which is $1 + 2\delta
R \cos(\omega R)/\sin(\omega R)$ to first order in $\delta R$.
%following:
%
%\begin{equation}
%sin^2((\omega + \delta)R)/sin^2(\omega R)= 1 + 2\delta R
%cos(\omega R)/sin(\omega R) - (\delta R)^2+...
%\end{equation}

The value of $\delta R$ here is $\sim 1\times 10^{-6}$ so the
assumption that the AOM has approximately the same effect on optical
photons differing by $\Delta$ is reasonable - i.e. the error is
negligible. One must further note that it has been assumed that all
photons are monochromatic: in reality they have a bandwidth. The AOM
has a frequency dependent action but providing that the bandwidth of
each photon is much less than their mean frequency the above
analysis can be applied again. Since the bandwidth is less than the
biexciton shift by at least an order of magnitude, this error is
negligible.

\section{Conclusion}
In this communication we showed that the inevitable irregularities
present in arrays of solid state cavities and the asymmetries of
individual quantum dots need not proscribe their use as sources for
entanglement generation or distribution. The frequency beam splitter
described opens up a window for frequency errors which can be much
larger than the bandwidth of individual photons. If physical
differences between two mechanisms yield photons which differ in
frequency greater than their individual bandwidth, this frequency
which-way information can prevent the detection of entanglement in
other degrees of freedom. We show how this which-way information can
sometimes be removed. Our scheme has, we believe, the twin virtues
of simplicity and general applicability and hope it will be useful
in scenarios beyond the examples described here.

\section{Acknowledgments}
Thanks to Sean Barrett, John Rarity, Tony Short and Andrew Shields.
NJ would like to thank the EPSRC and Toshiba Research Europe Ltd.
for support. TMS  acknowledges the QIP IRC for financial support.


\begin{thebibliography}{30}
\bibitem{note1} This coherent superposition is only conjectured.
It is possible that the two $x$ and two $y$ photons are emitted
incoherently - perhaps if $\nu$ oscillates unpredictably. Indirect
support for coherent emissions from similar systems has recently
been found \cite{edamatsu}
\end{thebibliography}
\end{document}